\documentclass[12pt]{article}
\pdfoutput=1 
\usepackage{amsmath,amssymb,amsthm}
\usepackage{bbm,latexsym}
\usepackage{graphicx}
\usepackage[numbers,sort&compress]{natbib}
\usepackage{authblk}
\usepackage{xcolor}
\newcommand{\zed}{\mathbbm{Z}}
\newcommand{\three}{\textbf{\underline{3}} }

\begin{document}
\title{
\LARGE \bf Direct and Semi-Direct Approaches to Lepton Mixing with a Massless Neutrino}
\setcounter{footnote}{2}
\author{Stephen F.\ King\thanks{E-mail: king@soton.ac.uk}\quad}
\author{\quad Patrick Otto Ludl\thanks{E-mail: P.Ludl@soton.ac.uk}}
\affil{{\small School of Physics and Astronomy, University of Southampton,\\Southampton, SO17 1BJ, United Kingdom}}

\date{June 24, 2016}

\maketitle

\begin{abstract}
We discuss the possibility of enforcing a massless Majorana neutrino in the direct and semi-direct approaches
to lepton mixing, in which the PMNS matrix is partly predicted by subgroups of a discrete family symmetry,
extending previous group searches up to order 1535.
We find a phenomenologically viable scheme for the semi-direct approach based on $Q(648)$ which
contains $\Delta(27)$ and the quaternion group as subgroups.
This leads to novel predictions for the first column of the PMNS matrix corresponding to a normal neutrino mass hierarchy with $m_1=0$, 
and sum rules for the mixing angles and phase which are characterised by the solar angle
being on the low side $\theta_{12}\sim 31^{\circ}$ and the Dirac (oscillation) CP phase 
$\delta$ being either about $\pm 45^\circ$ or $\pm \pi$.
\end{abstract}

\section{Introduction}

Neutrino mass and lepton mixing differs markedly from that of quarks in several ways. The extreme smallness
of neutrino mass, together with large lepton mixing provide fascinating clues in the search for a theory of flavour.
One idea is that lepton mixing may be governed by a discrete family symmetry group $G_f$
which controls the Majorana neutrino and charged lepton mass matrices leading to lepton mixing predictions \cite{King:2013eh,King:2014nza}. 
The three possible implementations of flavour symmetries are known as ``direct'', ``semi-direct'' and ``indirect''~\cite{King:2013eh,King:2014nza}. 

According to the ``direct'' approach, the Klein symmetry $G_{\nu}=\zed_2\times \zed_2$ of the 
Majorana neutrino mass matrix and the symmetry $G_\ell$ which fixes the form of the lepton mass matrix
are both subgroups of ${G}_{f}$, resulting in a prediction for all the lepton mixing angles and Dirac phase. 
The advantage of the direct approach is that the prediction arises purely from symmetry and does not require
any detailed knowledge of the model.
However, the direct approach requires a rather large
group~\cite{Holthausen:2012wt, King:2013vna,Fonseca:2014koa,Yao:2015dwa}, and the only
viable mixing pattern is the trimaximal mixing, with $\delta$ being either zero or $\pm \pi$. 

In the ``semi-direct'' approach, the symmetry of the neutrino mass matrix is typically reduced to $\zed_2$ for Majorana neutrinos, which constrains only the second column of the PMNS matrix to be $(1,1,1)^{T}/\sqrt{3}$, 
or the first column to be $(2,1,1)^{T}/\sqrt{6}$ (up to phases),
and the reactor angle $\theta_{13}$ can be accommodated with a small discrete group such as $S_4$. In the ``indirect'' approach, the flavour symmetry is completely broken such that the observed neutrino flavour symmetry emerges indirectly as an accidental symmetry, and the predictions are model dependent
(for a recent review see \cite{King:2015aea}).

The above direct and semi-direct approaches 
usually assume three non-degenerate Majorana neutrino masses.
There has recently been some discussion of how this picture changes if one of the Majorana neutrino masses
is zero \cite{Joshipura:2013pga,Joshipura:2014pqa}. In this case the phase of the massless neutrino field
is undetermined resulting in one of the $\zed_2$ factors being replaced by $\zed_n$, with the consequence that 
the determinant of the family symmetry ${G}_{f}$ 
need not be $\pm 1$, \textit{i.e.}\ it is a subgroup of $U(3)$ rather than $SU(3)$.
Although this opens up the possibility that a new type of viable direct model being found,
in fact only a no-go theorem results from such searches up to order 511 
\cite{Joshipura:2013pga,Joshipura:2014pqa}.

In the present paper we extend the reach of such searches for direct models with one massless neutrino
up to order 1535, but without phenomenological success. On the other hand we also perform
a new type of search for semi-direct models up to order 1535, and find a successful example of this kind,
based on the group, 
\begin{equation}
G_f \simeq (\Delta (27) \rtimes Q_8) \rtimes \zed_3 
\end{equation}
where $Q_8$ denotes the quaternion group of order 8
and the group is therefore of order $3^4\times 8 = 648$.
We denote this group as $Q(648)$.
This leads to a successful prediction for the first column of the PMNS matrix,
\begin{equation}
U_{\rm PMNS}=
\begin{pmatrix}
U_{e1} & - & -\\
U_{\mu 1} & -  & -\\
U_{\tau 1} & - & -
\end{pmatrix}
\end{equation}
corresponding to a normal neutrino mass hierarchy with $m_1=0$ and 
sum rules for the mixing angles and phase which are characterised by the solar angle
being on the low side $\theta_{12}\sim 31^{\circ}$ and the Dirac (oscillation) CP phase 
$\delta$ either about $\pm 45^\circ$ or $\pm \pi$ (e.g. the recent hint of $\delta \sim -\pi/2$ is not allowed).

Before we outline the details of our analysis, we would like to briefly
comment on dynamical settings enforcing $m_1=0$. The most prominent scenarios
of this kind are type-I seesaw models with two right-handed
neutrinos~\cite{King:1999mb}. Such models necessarily imply the mass
of the lightest neutrino to vanish, independent of any imposed flavour symmetries.
In the same way, the discrete residual symmetries discussed in this paper lead to
$m_1=0$ independent of the number of right-handed neutrino fields. Therefore, although
not directly connected, the two approaches can easily be combined.
For example if the three left-handed lepton doublets $L$ transform under a triplet
representation~$\textbf{\underline{3}}$
of the flavour group $G_f$ and three Higgs-doublets
(or alternatively three flavons) $\phi$ also transform under a triplet representation $\textbf{\underline{3}}'$
of $G_f$, and the tensor product $\textbf{\underline{3}}^\ast \otimes \textbf{\underline{3}}'$
contains one or two-dimensional representations, a Dirac Yukawa coupling $\overline{L}\,{\phi}\,\nu_R$
to two right-handed neutrinos is compatible with $G_f$.

The remainder of the paper is laid out as follows. In section 2 we review residual symmetries with a massless
Majorana neutrino and describe our strategy and results of group searches for direct and indirect models, for groups up to order 1535. In section 3 we present the results of a numerical phenomenological analysis,
and show that while the direct models are excluded by current data, there is a unique group (up to this order) 
which yields acceptable results in the semi-direct approach, leading to mixing sum rules and phenomenological predictions. Section 4 concludes the paper.

\section{Group searches with a massless neutrino}

\subsection{Residual symmetries in the lepton mass matrices}

The main question around which the framework of residual
symmetries in the fermion mass
matrices~\cite{lam1,lam2,grimus,Ge:2011ih,toorop,Ge:2011qn,hernandez,Holthausen:2012wt,Fonseca:2014koa}
has been constructed is
the question: \textit{When do symmetries of a $3\times 3$-matrix
(partly) fix the matrix which diagonalises it?}

The answer to this question is fairly simple. Consider unitary matrices
$S_i$ and $T_j$ which leave a complex symmetric matrix $M$ or a Hermitian
matrix $H$ invariant,\footnote{For application to the fermion mass problem
the study of Hermitian and complex symmetric matrices is sufficient. For the
lepton sector we will later have $M=M_\nu$ and $H=M_\ell M_\ell^\dagger$.} \textit{i.e.}\
\begin{subequations}
\begin{align}
& S_i^T M S_i = M \quad\forall i,\label{symmetries_nu}\\
& T_j^\dagger H T_j = H \quad\forall j.
\end{align}
\end{subequations}
The set of all matrices $S_i$ forms a group $G_M$ of unitary $3\times 3$-matrices.
In the same way, also the matrices $T_j$ form a unitary group $G_H$.
If the matrices $M$ and $H$ have non-degenerate singular values\footnote{
The singular values are the elements of the diagonalised matrix. This condition is
necessarily fulfilled in all relevant applications, because the singular values of $H$ will
be identified with the charged-lepton
masses squared and the singular values of $M$ will be identified with the three light-neutrino masses.} the
groups $G_M$ and $G_H$ must be \textit{Abelian}. For this case one can show that the
matrices $U_M$ and $U_H$ which (simultaneously) bring all $S_i$ and $T_j$ to diagonal
form, \textit{i.e.}\
\begin{subequations}
\begin{align}
& U_M^\dagger S_i U_M = \hat{S}_i \quad \forall i,\label{diag_S}\\
& U_H^\dagger T_j U_H = \hat{T}_j \quad \forall j,
\end{align}
\end{subequations}
also diagonalise $H$ and $M$ via\footnote{Let us, as an illustration, give the derivation of this fact for a complex symmetric
matrix $M$. From equation~(\ref{diag_M}) we have $M u_k = m_k u_k^\ast$ (no summation over $k$) for the columns
$u_k$ of $U_M$, \textit{i.e.}\ the $u_k$ are singular vectors of $M$ with singular values $m_k$.
From equation~(\ref{symmetries_nu}) we then have
\begin{equation}
S_i^T M S_i u_k = m_k u_k^\ast \Rightarrow M(S_i u_k) = m_k (S_i u_k)^\ast.
\end{equation}
Since the singular values $m_i$ are assumed to be non-degenerate, the corresponding singular vectors
are unique up to multiplication with a constant, \textit{i.e.}\ $S_i u_k = s_i u_k$. Therefore, the
$u_k$ are simultaneous eigenvectors to all $S_i$, which implies equation~(\ref{diag_S}).}~\cite{lam2}
\begin{subequations}
\begin{align}
& U_M^T M U_M = \hat{M},\label{diag_M}\\
& U_H^\dagger H U_H = \hat{H},
\end{align}
\end{subequations}
where hatted matrices are diagonal matrices. 
In other words \textit{the set of symmetries $S_i$ of a matrix $M$ fixes its diagonalising matrix $U_M$.}
The same statement holds for
$T_j$ and $H$.

Let us now apply these mathematical considerations to the case of residual symmetries in the lepton sector. The assumption of residual symmetries is
that there is a flavour symmetry group $G_f$ (acting on the lepton fields) which is spontaneously broken
to a symmetry group $G_\nu$ in the neutrino sector,
\begin{equation}
S_i^T M_\nu S_i = M_\nu, \quad S_i \in G_\nu,
\end{equation}
and a symmetry group $G_\ell$ in the charged-lepton sector,
\begin{equation}
T_j^\dagger M_\ell M_\ell^\dagger T_j = M_\ell M_\ell^\dagger, \quad T_j \in G_\ell.
\end{equation}
Identifying $M=M_\nu$ and $H=M_\ell M_\ell^\dagger$ in the above discussion,
we immediately find $G_\nu \subset G_M$ and $G_\ell \subset G_H$.
Therefore, the symmetry groups $G_\ell$ and $G_\nu$ potentially (but not necessarily)
impose constraints on the matrices $U_\nu$ and $U_\ell$, which diagonalise $M_\nu$
and $M_\ell M_\ell^\dagger$ via
\begin{align}
& U_\nu^T M_\nu U_\nu = \mathrm{diag}(m_1,m_2,m_3),\\
& U_\ell^\dagger M_\ell M_\ell^\dagger U_\ell = \mathrm{diag}(m_e^2,m_\mu^2,m_\tau^2),
\end{align}
and consequently also on the mixing matrix $U_\mathrm{PMNS} = U_\ell^\dagger U_\nu$.

In order to know how $G_\nu$ and $G_\ell$ constrain $U_\nu$ and $U_\ell$
we have to understand to which degree an Abelian unitary $3\times 3-$matrix group $A$ determines
the unitary matrix $U$ which simultaneously diagonalises all elements of $A$.
There are only three possibilities:
\begin{itemize}
 \item None of the common eigenvectors of the elements of $A$ is unique.\footnote{For eigenvectors \textit{unique}
here means unique up to multiplication with a complex number.} $\Rightarrow A=\{\mathbbm{1}_3\}$ and $U$ is
an arbitrary unitary matrix. 
 \item One of the common eigenvectors of the elements of $A$ is unique. $\Rightarrow$ One column of
$U$ is proportional to this eigenvector. $\Rightarrow$ One column of $U$ is fixed by $A$ up to a rephasing.
 \item All three of the common eigenvectors of the elements of $A$ are unique. $\Rightarrow$ Each column of
$U$ is proportional to one of these eigenvectors. $\Rightarrow$ $U$ is fixed by $A$ up to rephasing and
reordering of its columns.
\end{itemize}
According to this finding, one classifies models based on residual symmetries
into two categories:
\begin{itemize}
 \item[(A)] \underline{Direct models:} $G_\nu$ fixes $U_\nu$ and $G_\ell$ fixes $U_\ell$ (up to reordering and rephasing of
the columns). $\Rightarrow$ $U_\mathrm{PMNS}$ is fixed (up to reordering and rephasing of
rows and columns).
 \item[(B)] \underline{Semidirect models:} In one sector the full diagonalising matrix is fixed, in the other sector
only a column is fixed.
  \begin{itemize}
   \item[(B1)] $G_\ell$ fixes $U_\ell$, $G_\nu$ fixes a column of $U_\nu$. $\Rightarrow$ One column of
  $U_\mathrm{PMNS}$ is fixed up to permutation of its elements. One may choose (if not determined by a
  concrete model) which column of $U_\mathrm{PMNS}$ is fixed.
  \item[(B2)] $G_\ell$ fixes one column of $U_\ell$, $G_\nu$ fixes $U_\nu$. $\Rightarrow$ One row of
  $U_\mathrm{PMNS}$ is fixed up to permutation of its elements. One may choose (if not determined by a
   concrete model) which row of $U_\mathrm{PMNS}$ is fixed.
  \end{itemize}
\end{itemize}
In this work we will only study the cases A and B1.
Case B2 has for example been studied in~\cite{B2}.
One could in principle also consider a third very weakly restrictive case C for
which in each sector only one column is fixed (up to rephasing).
In this case only one element of $U_\mathrm{PMNS}$ would be fixed.
Due to its low predictive power, this scenario is usually not studied,
and also we will not study it here.

\subsection{Residual symmetries enforcing one massless neutrino}

The residual symmetry groups $G_\nu$ and $G_\ell$
are Abelian groups of unitary $3\times 3$-matrices. Therefore,
they are subgroups of $U(1)\times U(1)\times U(1)$. For the charged lepton
sector, this is also the maximal symmetry group,
\textit{i.e.}\ $G_H = U(1)\times U(1)\times U(1)$.
For Majorana neutrinos the situation is different. If all neutrinos are
massive, the maximal symmetry group is $G_M = \zed_2 \times \zed_2 \times \zed_2$,
while, if one neutrino is massless, also $G_M = U(1) \times \zed_2 \times \zed_2$
is allowed. The case of one massless neutrino has been studied in~\cite{Joshipura:2013pga,Joshipura:2014pqa}.

Here we will further elaborate on the case of a massless neutrino
within the framework of residual symmetries. The case of direct models
has been studied in~\cite{Joshipura:2014pqa} for all suitable finite groups
up to order 511. In the present paper we extend this analysis to order 1535.
Moreover, we discuss semidirect models (of type B1) with a massless neutrino,
also up to order 1535. Before we discuss the details of the group searches we
have performed, we want to have a look on the generic requirements
potential flavour groups $G_f$ with a massless neutrino have to fulfill. As outlined in detail
in appendix~\ref{appA}, viable groups $G_f$
\begin{itemize}
 \item must possess a faithful three-dimensional irreducible representation,
 \item must not be of the form $G_f \simeq G_f' \times \zed_n$ $(n>1)$ and
 \item must not be of the form of the
following
theorem by Joshipura and Patel~\cite{Joshipura:2014pqa}:
Let $G$ be a group of $3\times 3$ matrices which contains only elements of the form
``\textit{diagonal matrix of phases times permutation matrix}'', where the six permutation
matrices are given by
\begin{equation}
\begin{split}
& P_1  = \begin{pmatrix}
1 & 0 & 0\\
0 & 1 & 0\\
0 & 0 & 1
\end{pmatrix},\quad
P_2 = \begin{pmatrix}
0 & 1 & 0\\
1 & 0 & 0\\
0 & 0 & 1
\end{pmatrix},\quad
P_3 = \begin{pmatrix}
0 & 0 & 1\\
0 & 1 & 0\\
1 & 0 & 0
\end{pmatrix},\\
& P_4 = \begin{pmatrix}
1 & 0 & 0\\
0 & 0 & 1\\
0 & 1 & 0
\end{pmatrix},\quad
P_5 = \begin{pmatrix}
0 & 0 & 1\\
1 & 0 & 0\\
0 & 1 & 0
\end{pmatrix},\quad
P_6 = \begin{pmatrix}
0 & 1 & 0\\
0 & 0 & 1\\
1 & 0 & 0
\end{pmatrix}.
\end{split}
\end{equation}
Then, if such a group $G$ is used to build models enforcing a massless neutrino, the column vector
of the mixing matrix associated to the massless neutrino must be
\begin{equation}
\begin{pmatrix}
1\\0\\0
\end{pmatrix},\;
\frac{1}{\sqrt{3}}
\begin{pmatrix}
1\\1\\1
\end{pmatrix},\;
\frac{1}{\sqrt{2}}
\begin{pmatrix}
0\\1\\1
\end{pmatrix}
\end{equation}
or permutations thereof (\textit{i.e.}\ permutations
of the elements of an individual column.)
\end{itemize}
As a starting point for our analysis, we need a list of groups
fulfilling these criteria. We used the library
SmallGroups\footnote{Throughout the paper we will use the SmallGroups ID to identify groups.
This ID consists of two numbers in square
brackets, \textit{i.e.}\ $[g,n]$,
$g$ being the group order and $n$ being a label.
Two groups with different SmallGroups IDs are non-isomorphic.}~\cite{SmallGroups1,SmallGroups2}
and the computer algebra system GAP~\cite{GAP} to find all groups
of order smaller than 1536 which fulfill these minimal criteria.\footnote{Since we require
groups which have a three-dimensional irreducible representation, the group order
must be divisible by 3. Up to order 1535 there are 1342632 groups whose order is
divisible by three. For the group order 1536 alone there are 408641062 groups.
Therefore, we had to stop our searches at order 1535.} As a result of this scan we
have found 22 groups of order smaller than 1536 fulfilling the minimal criteria.\footnote{In this
paper we follow the approach of scanning over a set of eligible groups, in the end discarding
which are incompatible with experiment. The opposite approach of constructing eligible groups
directly from experimental data on the mixing matrices has \textit{e.g.}\ been used in~\cite{Grimus:2013rw,Talbert:2014bda}.}
They are shown in equation~(\ref{groups-massless}) in appendix~\ref{appA}.

\subsubsection{Direct models}

Let us now investigate the requirements for direct models
enforcing a massless neutrino. The requirement for the
charged-lepton sector is the usual one: Any group $G_\ell$
which uniquely determines the diagonalising matrix $U_\ell$
is sufficient. The same also holds for semidirect models of type~B1.
In the neutrino sector, we require a residual symmetry group
$G_\nu$ which completely fixes $U_\nu$ (as always up to rephasing and
reordering of the columns) and which enforces one neutrino mass
to vanish. The requirement of a vanishing neutrino mass implies
that $G_\nu$ is a subgroup of $U(1)\times \zed_2\times \zed_2$
instead of $\zed_2 \times \zed_2\times \zed_2$.
Therefore, there exists a basis in which all elements
of $G_\nu$ have the form
\begin{equation}
\begin{pmatrix}
\lambda & 0 & 0\\
0 & \alpha & 0\\
0 & 0 & \beta
\end{pmatrix}
\end{equation}
with $\lambda \in U(1)$, $\alpha,\beta\in \{-1,+1\}$.
Moreover, in order to enforce a vanishing neutrino mass,
for at least one element of $G_\nu$ we must have $\lambda\neq\pm 1$.
This element may be of four forms:
\begin{equation}
\begin{split}
& S_1 = \mathrm{diag}(\lambda,+1,+1),\quad
  S_2 = \mathrm{diag}(\lambda,+1,-1),\\
& S_3 = \mathrm{diag}(\lambda,-1,+1),\quad
  S_4 = \mathrm{diag}(\lambda,-1,-1).
\end{split}
\end{equation}
Every group $G_\nu$ capable of fixing the complete matrix $U_\nu$
and enforcing a massless neutrino
contains at least one element of the form $S_2$ or $S_3$. Namely, if it did
not contain such an element, in order to fulfill all requirements it
would have to contain at least one element of the form
\begin{equation}\label{Seq1}
\mathrm{diag}(\lambda,\pm1,\pm1)
\end{equation}
and one further element with non-degenerate 22 and 33 elements, \textit{i.e.}\
\begin{equation}\label{Seq2}
\mathrm{diag}(\pm1,\pm1,\mp1) \quad\text{or}\quad
\mathrm{diag}(\pm1,\mp1,\pm1).
\end{equation}
However, the product of the matrices of equations~(\ref{Seq1})
and~(\ref{Seq2}) is of the form $S_2$ or $S_3$, which proves
that $G_\nu$ always contains an element of this form.
The matrices $S_2$ and $S_3$, since they have non-degenerate
eigenvalues, on their own already fix the complete matrix $U_\nu$.
Therefore, we can restrict the
analysis to groups $G_\nu'=\langle\langle S_2 \rangle\rangle \subset G_\nu$ generated by $S_2$ (or
$G_\nu'=\langle\langle S_3 \rangle\rangle \subset G_\nu$ generated by $S_3$.)\footnote{In this paper
the symbol $\langle\langle\ldots\rangle\rangle$ means ``generated by $\ldots$''.}
Thus, the requirement on $G_f$ is:
\begin{itemize}
 \item $G_f$ has a faithful three-dimensional irreducible representation
which has at least one element with one eigenvalue $\lambda\neq \pm 1$,
one eigenvalue $+1$ and one eigenvalue $-1$.
This element generates the residual symmetry group $G_\nu' \simeq \zed_n$ ($n$ even)
and determines $U_\nu$ up to rephasing and reordering of the columns.
\end{itemize}
Thus, among the groups of equation~(\ref{groups-massless}) we search for those which have a faithful three-dimensional irreducible
representation \three (defining a matrix group $\three(G_f)$ isomorphic to $G_f$) fulfilling the following criteria:
\begin{itemize}
 \item $\three(G_f)$ contains a matrix $S$ with eigenvalues $\{\lambda,+1,-1\}$, $\lambda\neq \pm 1$.
This is a basis independent property and may easily be checked by testing $\mathrm{Tr}\,S+\mathrm{det}\,S=0$,
$\mathrm{Tr}\,S\neq \pm 1$. The matrix $S$ is then a candidate for a generator of $G_\nu'$.
 \item $\three(G_f)$ contains an Abelian subgroup $G_\ell$ which can completely fix its diagonalising matrix $U_\ell$.
 \item There must be choices of $G_\ell'$ and $G_\nu'$ such that
$G_\nu'$ is not a subgroup of $G_\ell'$ and vice versa. (Otherwise the mixing matrix $U_\mathrm{PMNS}$
would be trivial!) Also, none of the generators of $G_\ell'$ must commute with the generator
$S$ of $G_\nu'$---see appendix~\ref{appB}.
 \item Moreover, the groups $G_\ell'$ and $G_\nu'$ must together generate the whole matrix group
$\three(G_f)$. Namely, if they do not, we can restrict ourselves to the subgroup
$G_f'\equiv \langle\langle G_\ell', G_\nu' \rangle\rangle \subset \three(G_f)$. If $G_f'$
is an irreducible matrix group, $G_f'$ fulfills all criteria for our search and predicts
the same mixing matrix as $\three(G_f)$. If $\three(G_f')$ is reducible, the mixing matrix
has two vanishing mixing angles and is therefore not compatible with experiment.
\end{itemize}
Performing a group search with GAP, we find that only seven groups
of order smaller than 1536 fulfill all these criteria:
\begin{equation}
[ 108, 15 ], [ 324, 111 ], [ 432, 239 ], [ 648, 533 ], [ 864, 675 ], [ 972, 411 ], [ 1296, 1995 ].
\end{equation}

\subsubsection{Semidirect models}

The requirements for semidirect models of type~(B1) are identical with only two
differences:
\begin{itemize}
 \item The requirement on $S$ now becomes: 
$\three(G_f)$ contains a matrix $S$ with eigenvalues $\{\lambda,\pm1,\pm1\}$, $\lambda\neq \pm 1$,
respectively (\textit{i.e.}\ two degenerate eigenvalues in each case).
This ensures that one mass is set to zero, but only one column of $U_\nu$ is fixed.
 \item The argument used to prove that none of the generators of $G_\ell$
must commute with the generator $S$ of $G_\nu$---see appendix~\ref{appB}---does not
hold for semidirect models. Thus, one of the generators of
$G_\ell'$ is allowed to commute with $S$.
\end{itemize}
Also for the semidirect case the groups $G_\ell'$ and $G_\nu'$ have to generate
the full group $\three(G_f)$. However, the argument for this is different to the
case of direct models. Again, if $\three(G_f')$ is irreducible, we may replace $G_f$
by $\three(G_f')$. If $\three(G_f')$ is reducible one can (by the same argument
as used in appendix~\ref{appA} to show that $\three(G_f)$ must be irreducible)
show that a reducible $\three(G_f')$ leads to predictions incompatible with
experiment.

Doing a group search with GAP, one finds that there is only one single group
of order smaller than 1536 which meets all requirements for a semidirect model.
This group has the identification number $[648,533]$ in the SmallGroups library.

\section{Numerical analysis and phenomenology}

\subsection{Direct models}

In order to test the seven candidate groups for direct models
with a massless neutrino, we computed all faithful three-dimensional
irreducible representations of the groups, computed all Abelian subgroups
and listed all possible combinations $(G_\ell,G_\nu)$.
For each of these combinations, the possible mixing matrices
have been computed---see~\cite{Holthausen:2012wt} for a detailed description
of this procedure. 
In order to compare the predictions for the mixing matrix
with experiment we fitted the three mixing angles to the
global fit data of~\cite{nufit}. As $\chi^2$-function we used
\begin{equation}
\chi^2(\theta_{12},\,\theta_{23},\,\theta_{13}) \equiv
\sum_{ij=12,23,13} \left(\frac{\mathrm{sin}^2\theta_{ij}^\text{exp}-\mathrm{sin}^2\theta_{ij}^\text{pred}}{\sigma(\mathrm{sin}^2\theta_{ij})}\right)^2,
\end{equation}
which has three degrees of freedom. For the errors $\sigma(\mathrm{sin}^2\theta_{ij})$ we used the
values given in~\cite{nufit} (in case of an asymmetric error distribution we used the larger error).
The resulting minimal values of $\chi^2$ are listed in table~\ref{chisquaredtable}.
\begin{table}
\begin{center}
\begin{tabular}{|c|c|c|}
\hline
Group & $\chi^2_\mathrm{min}$ (normal spectrum) & $\chi^2_\mathrm{min}$ (inverted spectrum) \\
\hline
$[ 108, 15 ]$ & $1.80 \times 10^{2}$ & $4.52 \times 10^{2}$ \\
$[ 324, 111 ]$ & $1.80 \times 10^{2}$ & $4.52 \times 10^{2}$ \\
$[ 432, 239 ]$ & $1.07 \times 10^{4}$ & $1.56 \times 10^{4}$ \\
$[ 648, 533 ]$ & $1.30 \times 10^{2}$ & $4.19 \times 10^{2}$ \\
$[ 864, 675 ]$ & $1.07 \times 10^{4}$ & $1.56 \times 10^{4}$ \\
$[ 972, 411 ]$ & $1.80 \times 10^{2}$ & $4.52 \times 10^{2}$ \\
$[ 1296, 1995 ]$ & $1.07 \times 10^{4}$ & $1.56 \times 10^{4}$ \\
\hline
\end{tabular}
\end{center}
\caption{The minimal values of $\chi^2$ for the seven candidate groups
for direct models. Groups which lead to the same $\chi^2_\mathrm{min}$
predict the same values for the elements $|U_{ij}|$ of the
mixing matrix.}\label{chisquaredtable}
\end{table}
Evidently, none of the candidate groups is compatible with
the experimental data. For groups up to order 511 this result
has been found earlier in~\cite{Joshipura:2014pqa}.

\subsection{Semidirect models}

For the semidirect models, there is a unique candidate
group $[648,533]$. It has six faithful three-dimensional irreducible
representations, each of which can predict (the same) 19 different
patterns for a column of the mixing matrix.
It is therefore sufficient to study only one of the faithful
three-dimensional irreducible representations constructed with GAP,
\textit{i.e.}\ we pick one of them and use it to define the group
$[648,533]$ as a matrix group.
In this representation the group is generated by the two matrices
\begin{equation}
S = \frac{1}{3}
\begin{pmatrix}
1+\epsilon^4-\epsilon^6-\epsilon^8 & -\epsilon^2+\epsilon^7 & -\epsilon^2+\epsilon^4  \\
\epsilon-\epsilon^5 & 1+\epsilon-\epsilon^2-\epsilon^6 & \epsilon^4-\epsilon^5 \\
\epsilon-\epsilon^8 & \epsilon^7 - \epsilon^8 & 1-\epsilon^5-\epsilon^6+\epsilon^7
\end{pmatrix}
\end{equation}
and
\begin{equation}
T = \begin{pmatrix}
\epsilon^2 & 0 & 0 \\
0 & \epsilon^5 & 0 \\
0 & 0 & \epsilon^8
\end{pmatrix},
\end{equation}
where
\begin{equation}
\epsilon \equiv \exp(2\pi i/9),
\end{equation}
\textit{i.e.}\
\begin{equation}
G_f = \langle\langle S,\, T \rangle\rangle  \simeq (((\zed_3 \times \zed_3) \rtimes \zed_3) \rtimes Q_8) \rtimes \zed_3 \simeq (\Delta (27) \rtimes Q_8) \rtimes \zed_3 .
\end{equation}
$Q_8$ here denotes the quaternion group of order 8.
This group corresponds to $[648,533] $ which 
we denote as $Q(648)$.

The two only columns predictable by $Q(648)$
being compatible with experiment emerge from the choice
\begin{equation}
G_\nu = \langle\langle S \rangle\rangle \simeq \zed_3, \quad
G_\ell = \langle\langle T \rangle\rangle \simeq \zed_9.
\end{equation}
We are already in a basis where all elements of $G_\ell$ are diagonal.
Therefore, the eigenvector of $S$ with eigenvalue $\neq \pm 1$ is
the predicted column of $U_\mathrm{PMNS}$. Indeed, $S$ has two eigenvalues $+1$
and an eigenvalue $\omega \equiv \exp(2\pi i/3) = \epsilon^3$
with the corresponding eigenvector
\begin{equation}
u = \frac{1}{3}
\begin{pmatrix}
1 + \epsilon + \epsilon^8\\
\epsilon^4 + \epsilon^6 + \epsilon^8\\
\epsilon^3 + \epsilon^7 + \epsilon^8
\end{pmatrix}.
\end{equation}
The absolute values of the entries of this vector are
\begin{align}
& |u_1| = \frac{1}{3} \left( 1 + 2\,\mathrm{cos}\frac{2\pi}{9} \right) \approx 0.844,\\
& |u_2| = \frac{1}{3} \left( 1 + 2\,\mathrm{cos}\frac{4\pi}{9} \right) \approx 0.449,\\
& |u_3| =-\frac{1}{3} \left( 1 + 2\,\mathrm{cos}\frac{8\pi}{9} \right) \approx 0.293.
\end{align}
\begin{figure}
\begin{center}
\includegraphics[width=0.5\textwidth]{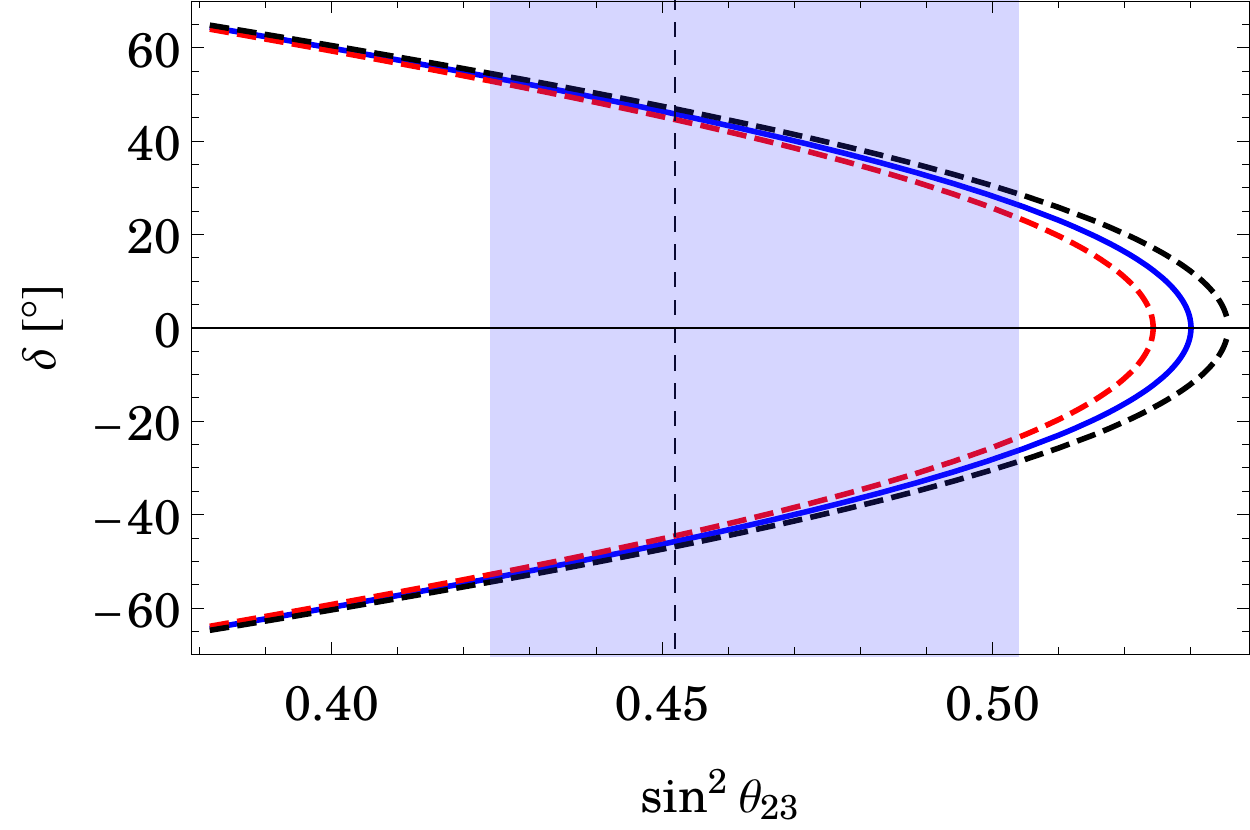}\hspace*{5mm}
\includegraphics[width=0.5\textwidth]{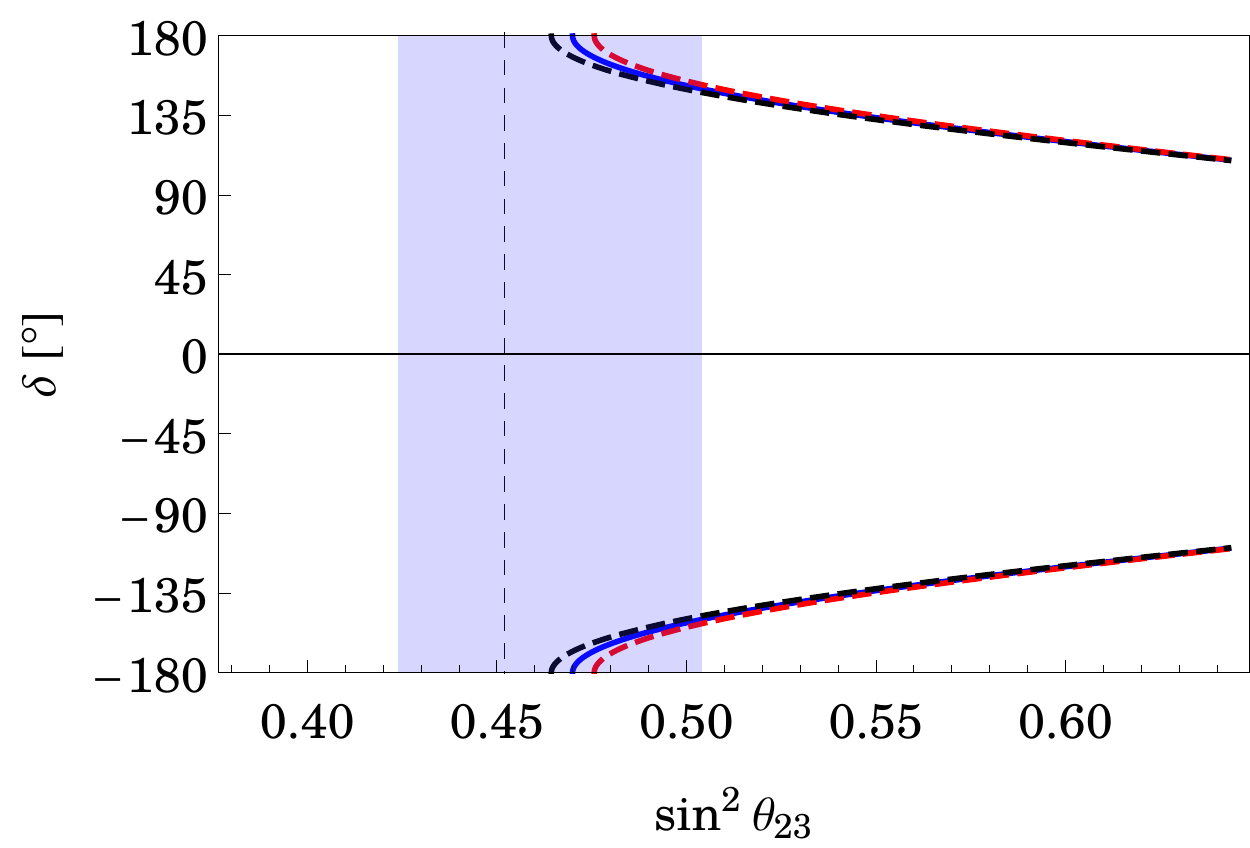}\\
\end{center}
\caption{The value of the Dirac phase $\delta$ as a function of $\mathrm{sin}^2\theta_{23}$
for different values of $\mathrm{sin}^2\theta_{13}$ according to the global fit of~\cite{nufit}:
best-fit (blue), $1\sigma$ lower bound (red dashed), $1\sigma$ upper bound (black dashed). The dashed vertical
line indicates the best-fit value $\mathrm{sin}^2\theta_{23}=0.452$ and the blue shaded area is the $1\sigma$
region for $\mathrm{sin}^2\theta_{23}$.
Left plot: $(|U_{e1}|,|U_{\mu 1}|)=(|u_1|,|u_2|)$, right plot: $(|U_{e1}|,|U_{\mu 1}|)=(|u_1|,|u_3|)$.}\label{plots}
\end{figure}
This leads to two patterns compatible with the first column\footnote{Since the neutrino
mass associated with the predicted column vanishes, we have $m_1=0$ which requires
a normal neutrino mass spectrum.} of $U_\mathrm{PMNS}$,
namely:
\begin{equation}
\begin{pmatrix}
|U_{e1}| \\
|U_{\mu 1}| \\
|U_{\tau 1}|
\end{pmatrix} =
\begin{pmatrix}
|u_1| \\
|u_2| \\
|u_3|
\end{pmatrix}
\quad
\text{and}
\quad
\begin{pmatrix}
|U_{e1}| \\
|U_{\mu 1}| \\
|U_{\tau 1}|
\end{pmatrix} =
\begin{pmatrix}
|u_1| \\
|u_3| \\
|u_2|
\end{pmatrix}.
\end{equation}
Note that we here have used the permutation freedom of the elements of
the predicted column, which comes from the fact that the
residual symmetries cannot fix any mass orderings.
The equations
\begin{subequations}
\begin{align}
& |U_{e1}| = c_{12}\,c_{13}, \\
& |U_{\mu 1}| = |s_{12} c_{23} + c_{12} s_{13} s_{23} e^{i\delta}|,
\end{align}
\end{subequations}
where $s_{ij}\equiv \mathrm{sin}\,\theta_{ij}$
and $c_{ij}\equiv \mathrm{cos}\,\theta_{ij}$,
give relations between the mixing angles and the
Dirac phase $\delta$.
These relations can be used to predict
$\theta_{12}$ as a function of $\theta_{13}$
and $\mathrm{cos}\,\delta$ as a function of
$\theta_{13}$ and $\theta_{23}$:
\begin{subequations}
\begin{align}
& s_{12}^2 = 1 - |U_{e1}|^2 (1+t_{13}^2),\\
& \mathrm{cos}\,\delta = \frac{|U_{\mu 1}|^2 \left(1+t_{23}^2\right)+|U_{e1}|^2 \left(t_{23}^2-\left(1+t_{13}^2\right) \left(t_{23}^2-1\right)\right)-1}{2 \, |U_{e1}| \, t_{13} \,
t_{23} \sqrt{1-|U_{e1}|^2 \left(1+t_{13}^2\right)}}.
\end{align}
\end{subequations}
Here $t_{ij}\equiv \mathrm{tan}\,\theta_{ij}$.
Using the values of the global fit of ref.~\cite{nufit},
\begin{equation}\label{best-fit}
\mathrm{sin}^2\theta_{23} = 0.452^{+0.052}_{-0.028}, \quad \mathrm{sin}^2\theta_{13} = 0.0218^{+0.0010}_{-0.0010},
\end{equation}
we find (best-fit)
\begin{subequations}
\begin{alignat}{2}
& \mathrm{sin}^2\theta_{12} = 0.272, \quad \theta_{12}=31.4^\circ, \\
& \mathrm{cos}\,\delta = 0.697, \quad \delta = \pm 45.8^\circ
\end{alignat}
\end{subequations}
for $|U_{e1}|=|u_1|$, $|U_{\mu 1}|=|u_2|$.
The other possibility with $|U_{e1}|=|u_1|$, $|U_{\mu 1}|=|u_3|$ 
turns out to be incompatible with the best-fit values
of equation~(\ref{best-fit}),
\begin{subequations}
\begin{alignat}{2}
& \mathrm{sin}^2\theta_{12} = 0.272, \quad \theta_{12}=31.4^\circ, \\
& \mathrm{cos}\,\delta = -1.07 \rightarrow \text{ inconsistent,}
\end{alignat}
\end{subequations}
but remains consistent with experiment for $s_{23}^2\gtrsim 0.47$.
Using the best-fit values of the mass-squared differences and $\mathrm{sin}^2\theta_{13}$
from~\cite{nufit} as input parameters,
the predicted range for $m_{\beta\beta}$ is $(1.22\div 3.38)\,\mathrm{meV}$, \textit{i.e.}\
several meV (as for every model with a normal neutrino mass spectrum and vanishing $m_1$).
Since $m_{\beta\beta}$ depends only on the sum of $\delta$ and one of the
(unconstrained) Majorana phases,
our model puts no stronger constraint on $m_{\beta\beta}$.

In total, the two discussed column patterns are compatible with the
global fit values of~\cite{nufit} at about $2-3$ sigma.
The reason for tension is the too small value of the solar
mixing angle predicted by the group using the reactor angle as an input.
The global-fit result for the solar mixing angle is
\begin{equation}
\mathrm{sin}^2\theta_{12} = 0.304^{+0.013}_{-0.012}, \quad \theta_{12} = (33.48^{+0.78}_{-0.75})\hspace{0mm}^\circ.
\end{equation}
The main prediction of the model is a value of $\delta$ of about
$\pm 45^\circ$ or $\pm \pi$ (for the best-fit values).
The value of $\delta$ for different values of $s_{23}^2$
and $s_{13}^2$ is shown in figure~\ref{plots}.

\section{Conclusions}

In this paper we have 
discussed the possibility of enforcing a massless Majorana neutrino in the direct and semi-direct approaches
to lepton mixing, in which the PMNS matrix is partly predicted by subgroups of a discrete family symmetry.
Our analysis extends previous group searches for direct models from order 511
up to 1535, and provides the first analysis of semi-direct models with a massless
neutrino up to this order. Our results confirm and extend the no-go results of Joshipura and Patel
up to order 1535 for the direct approach.

However, we find a new phenomenologically viable scheme for the semi-direct approach
based on $Q(648)$ which contains $\Delta(27)$ and the quaternion group as subgroups.
This leads to novel predictions for the first column of the PMNS matrix corresponding to a normal neutrino mass hierarchy with $m_1=0$, 
and sum rules for the mixing angles and phase which are characterised by the solar angle
being on the low side $\theta_{12}\sim 31^{\circ}$ and the Dirac (oscillation) CP phase 
$\delta$ being either about $\pm 45^\circ$ or $\pm \pi$.

\paragraph{Acknowledgements:}
The authors acknowledge support from the STFC grant\\
ST/L000296/1 and 
the European Union Horizon 2020 research and innovation programme under the Marie Sklodowska-Curie grant agreements
InvisiblesPlus RISE No.\ 690575 and Elusives ITN No.\ 674896.

\begin{appendix}

\section{Generic requirements on flavour symmetries in the
framework of residual symmetries with one massless neutrino.}\label{appA}

In order to be potentially phenomenologically viable
in the framework of residual symmetries
in the lepton sector, a flavour group
$G_f$ must fulfill two generic requirements:

\begin{itemize}
 \item $G_f$ must have a \textit{faithful three-dimensional irreducible representation}.
The requirement for a three-dimensional representation comes from the fact
that there are three generations of leptons. This representation must be faithful because
otherwise we could restrict ourselves to the smaller group defined by the non-faithful matrix
representation. Moreover, the three-dimensional faithful representation under consideration
must also be irreducible. Namely, if it was reducible, there would be a basis in which
the matrices of $G_\nu$ and $G_\ell$ are simultaneously block-diagonal. In direct models
this would mean that also the mixing matrix is block-diagonal, thus implying two vanishing
mixing angles, which is clearly not compatible with experimental observations.
For the case of semidirect models
we consider the example of models of type B1.
The arguments for case B2 are analogous.
In the block-diagonal basis we have
\begin{equation}
G_\nu:
\begin{pmatrix}
\times & 0 & 0\\
0 & \times & \times\\
0 & \times & \times
\end{pmatrix},\ldots
\quad
G_\ell:
\begin{pmatrix}
\times & 0 & 0\\
0 & \times & \times\\
0 & \times & \times
\end{pmatrix},\ldots,
\end{equation}
where $\times$ stands for a non-zero entry.
We can now make a further basis transformation (a unitary 23-rotation in our example)
which makes all elements of $G_\ell$ diagonal, \textit{i.e.}\
\begin{equation}
G_\nu:
\begin{pmatrix}
\times & 0 & 0\\
0 & \times & \times\\
0 & \times & \times
\end{pmatrix},\ldots
\quad
G_\ell:
\begin{pmatrix}
\times & 0 & 0\\
0 & \times & 0\\
0 & 0 & \times
\end{pmatrix},\ldots.
\end{equation}
In this basis, the column of $U_\mathrm{PMNS}$ which is fixed by the semidirect
model is a common eigenvector of the matrices of $G_\nu$. But all matrices
of $G_\nu$ are still block-diagonal, which means that this common eigenvector
can only be of the form
\begin{equation}
\begin{pmatrix}
\times\\
0\\
0
\end{pmatrix}
\quad
\text{or}
\quad
\begin{pmatrix}
0\\
\times\\
\times
\end{pmatrix}.
\end{equation}
Thus, $U_\mathrm{PMNS}$ would contain at least one vanishing element,
which is phenomenologically not viable.
 \item We can \textit{discard all groups of the form $G_f = G_f' \times \zed_n$ $(n>1)$.}
Namely, since the relevant representation of $G_f$ must be
irreducible, the elements of $\zed_n$ are represented as matrices proportional
to $\mathbbm{1}_3$. Such symmetries cannot constrain the mixing matrix $U_\text{PMNS}$.
Therefore, it is sufficient to confine the study to the smaller group $G_f'$.
\end{itemize}
There are 384 groups of order smaller than 1536 which fulfill
these two criteria, they are shown in equation~(\ref{U3subgroups}).
This list extends the list of finite subgroups of U(3) found in~\cite{U3-512} to order 1535.

In the case of massless neutrinos
there is a third constraint:
\begin{itemize}
 \item 
Consider matrix groups which have only elements of the form
``\textit{diagonal matrix of phases times permutation matrix}'', where by
permutation matrices we mean the six matrices
\begin{equation}\label{permutationmatrices}
\begin{split}
& P_1  = \begin{pmatrix}
1 & 0 & 0\\
0 & 1 & 0\\
0 & 0 & 1
\end{pmatrix},\quad
P_2 = \begin{pmatrix}
0 & 1 & 0\\
1 & 0 & 0\\
0 & 0 & 1
\end{pmatrix},\quad
P_3 = \begin{pmatrix}
0 & 0 & 1\\
0 & 1 & 0\\
1 & 0 & 0
\end{pmatrix},\\
& P_4 = \begin{pmatrix}
1 & 0 & 0\\
0 & 0 & 1\\
0 & 1 & 0
\end{pmatrix},\quad
P_5 = \begin{pmatrix}
0 & 0 & 1\\
1 & 0 & 0\\
0 & 1 & 0
\end{pmatrix},\quad
P_6 = \begin{pmatrix}
0 & 1 & 0\\
0 & 0 & 1\\
1 & 0 & 0
\end{pmatrix}.
\end{split}
\end{equation}
It has been shown by Joshipura and Patel in~\cite{Joshipura:2014pqa}
that models based on such groups with a massless neutrino (enforced by the
residual symmetry) can only lead to the following columns of the
mixing matrix (absolute values of the entries
of the fixed column of $U_\text{PMNS}$)
associated to the massless neutrino:
\begin{equation}
\begin{pmatrix}
1\\0\\0
\end{pmatrix},\;
\frac{1}{\sqrt{3}}
\begin{pmatrix}
1\\1\\1
\end{pmatrix},\;
\frac{1}{\sqrt{2}}
\begin{pmatrix}
0\\1\\1
\end{pmatrix}
\end{equation}
and permutations thereof (\textit{i.e.}\ permutations
of the elements of an individual column.)
The only phenomenologically viable case here is
\begin{equation}
\frac{1}{\sqrt{3}}
\begin{pmatrix}
1\\1\\1
\end{pmatrix},
\end{equation}
which is called $\mathrm{TM}_2$ in the literature, since it
fits the second column of the lepton mixing matrix.
However, this scenario would predict $m_2=0$,
which is excluded by experiment.
Therefore, we can exclude also all groups which are of the form
discussed in the theorem by Joshipura and Patel.
\end{itemize}
There are only 22 groups of order smaller than 1536 which also fulfill
the third requirement:
\begin{equation}
\begin{split}\label{groups-massless}
& [ 60, 5 ], [ 108, 15 ], [ 168, 42 ], [ 216, 25 ], [ 216, 88 ], [ 324, 111 ], [ 432, 57 ], [ 432, 239 ], [ 432, 273 ],\\
& [ 648, 352 ], [ 648, 531 ], [ 648, 532 ], [ 648, 533 ], [ 648, 551 ], [ 864, 194 ], [ 864, 675 ], [ 864, 737 ],\\
& [ 972, 411 ], [ 1080, 260 ], [ 1296, 1239 ], [ 1296, 1995 ], [ 1296, 2203 ].
\end{split}
\end{equation}
Therefore, for the study of massless neutrinos in the framework
of residual symmetries, confining oneself to flavour symmetry
groups of order smaller than 1536, it is sufficient to study the
22 groups of equation~(\ref{groups-massless}).

\begin{footnotesize}
\begin{equation}\label{U3subgroups}
\begin{split}
& [ 12, 3 ], [ 21, 1 ], [ 24, 12 ], [ 27, 3 ], [ 27, 4 ], [ 36, 3 ], [ 39, 1 ], [ 48, 3 ], [ 48, 30 ], [ 54, 8 ],\\
& [ 57, 1 ], [ 60, 5 ], [ 63, 1 ], [ 75, 2 ], [ 81, 6 ], [ 81, 7 ], [ 81, 8 ], [ 81, 9 ], [ 81, 10 ], [ 81, 14 ], [ 84, 11 ],\\
& [ 93, 1 ], [ 96, 64 ], [ 96, 65 ], [ 108, 3 ], [ 108, 11 ], [ 108, 15 ], [ 108, 19 ], [ 108, 21 ], [ 108, 22 ], [ 111, 1 ],\\
& [ 117, 1 ], [ 129, 1 ], [ 144, 3 ], [ 147, 1 ], [ 147, 5 ], [ 150, 5 ], [ 156, 14 ], [ 162, 10 ], [ 162, 12 ], [ 162, 14 ],\\
& [ 162, 44 ], [ 168, 42 ], [ 171, 1 ], [ 183, 1 ], [ 189, 1 ], [ 189, 4 ], [ 189, 5 ], [ 189, 7 ], [ 189, 8 ],\\
& [ 192, 3 ], [ 192, 182 ], [ 192, 186 ], [ 201, 1 ], [ 216, 17 ], [ 216, 25 ], [ 216, 88 ], [ 216, 95 ], [ 219, 1 ],\\
& [ 225, 3 ], [ 228, 11 ], [ 237, 1 ], [ 243, 16 ], [ 243, 19 ], [ 243, 20 ], [ 243, 24 ], [ 243, 25 ], [ 243, 26 ],\\
& [ 243, 27 ], [ 243, 50 ], [ 243, 55 ], [ 252, 11 ], [ 273, 3 ], [ 273, 4 ], [ 279, 1 ], [ 291, 1 ], [ 294, 7 ], [ 300, 13 ],\\
& [ 300, 43 ], [ 309, 1 ], [ 324, 3 ], [ 324, 13 ], [ 324, 15 ], [ 324, 17 ], [ 324, 43 ], [ 324, 45 ], [ 324, 49 ],\\
& [ 324, 50 ], [ 324, 51 ], [ 324, 60 ], [ 324, 102 ], [ 324, 111 ], [ 324, 128 ], [ 327, 1 ], [ 333, 1 ], [ 336, 57 ],\\
& [ 351, 1 ], [ 351, 4 ], [ 351, 5 ], [ 351, 7 ], [ 351, 8 ], [ 363, 2 ], [ 372, 11 ], [ 381, 1 ], [ 384, 568 ], [ 384, 571 ],\\
& [ 384, 581 ], [ 387, 1 ], [ 399, 3 ], [ 399, 4 ], [ 417, 1 ], [ 432, 3 ], [ 432, 33 ], [ 432, 57 ], [ 432, 100 ],\\
& [ 432, 102 ], [ 432, 103 ], [ 432, 239 ], [ 432, 260 ], [ 432, 273 ], [ 441, 1 ], [ 441, 7 ], [ 444, 14 ], [ 453, 1 ],\\
& [ 468, 14 ], [ 471, 1 ], [ 486, 26 ], [ 486, 28 ], [ 486, 61 ], [ 486, 125 ], [ 486, 164 ], [ 489, 1 ], [ 507, 1 ],\\
& [ 507, 5 ], [ 513, 1 ], [ 513, 5 ], [ 513, 6 ], [ 513, 8 ], [ 513, 9 ], [ 516, 11 ], [ 525, 5 ], [ 543, 1 ], [ 549, 1 ],\\
& [ 567, 1 ], [ 567, 4 ], [ 567, 5 ], [ 567, 7 ], [ 567, 12 ], [ 567, 13 ], [ 567, 14 ], [ 567, 23 ], [ 567, 36 ], [ 576, 3 ],\\
& [ 579, 1 ], [ 588, 11 ], [ 588, 16 ], [ 588, 60 ], [ 597, 1 ], [ 600, 45 ], [ 600, 179 ], [ 603, 1 ], [ 624, 60 ],\\
& [ 633, 1 ], [ 648, 19 ], [ 648, 21 ], [ 648, 23 ], [ 648, 244 ], [ 648, 259 ], [ 648, 260 ], [ 648, 266 ], [ 648, 352 ],\\
& [ 648, 531 ], [ 648, 532 ], [ 648, 533 ], [ 648, 551 ], [ 648, 563 ], [ 651, 3 ], [ 651, 4 ], [ 657, 1 ], [ 669, 1 ],\\
& [ 675, 5 ], [ 675, 9 ], [ 675, 11 ], [ 675, 12 ], [ 684, 11 ], [ 687, 1 ], [ 711, 1 ], [ 723, 1 ], [ 726, 5 ], [ 729, 62 ],\\
& [ 729, 63 ], [ 729, 64 ], [ 729, 80 ], [ 729, 86 ], [ 729, 94 ], [ 729, 95 ], [ 729, 96 ], [ 729, 97 ], [ 729, 98 ],\\
& [ 729, 284 ], [ 729, 393 ], [ 729, 397 ], [ 732, 14 ], [ 741, 3 ], [ 741, 4 ], [ 756, 11 ], [ 756, 113 ], [ 756, 114 ],\\
& [ 756, 116 ], [ 756, 117 ], [ 768, 1083477 ], [ 768, 1085333 ], [ 768, 1085335 ], [ 768, 1085351 ], [ 777, 3 ], [ 777, 4 ],\\
& [ 804, 11 ], [ 813, 1 ], [ 819, 3 ], [ 819, 4 ], [ 831, 1 ], [ 837, 1 ], [ 837, 4 ], [ 837, 5 ], [ 837, 7 ], [ 837, 8 ],\\
& [ 849, 1 ], [ 864, 69 ], [ 864, 194 ], [ 864, 675 ], [ 864, 701 ], [ 864, 703 ], [ 864, 737 ], [ 867, 2 ], [ 873, 1 ],\\
& [ 876, 14 ], [ 900, 66 ], [ 903, 5 ], [ 903, 6 ], [ 912, 57 ], [ 921, 1 ], [ 927, 1 ], [ 939, 1 ], [ 948, 11 ], [ 972, 3 ],\\
& [ 972, 29 ], [ 972, 31 ], [ 972, 64 ], [ 972, 117 ], [ 972, 121 ], [ 972, 122 ], [ 972, 123 ], [ 972, 147 ], [ 972, 152 ],\\
& [ 972, 153 ], [ 972, 170 ], [ 972, 309 ], [ 972, 348 ], [ 972, 411 ], [ 972, 520 ], [ 972, 550 ], [ 975, 5 ], [ 981, 1 ],\\
& [ 993, 1 ], [ 999, 1 ], [ 999, 5 ], [ 999, 6 ], [ 999, 8 ], [ 999, 9 ], [ 1008, 57 ], [ 1011, 1 ], [ 1014, 7 ], [ 1029, 6 ],\\
& [ 1029, 9 ], [ 1047, 1 ], [ 1053, 16 ], [ 1053, 25 ], [ 1053, 26 ], [ 1053, 27 ], [ 1053, 29 ], [ 1053, 32 ], [ 1053, 35 ],\\
& [ 1053, 37 ], [ 1053, 47 ], [ 1080, 260 ], [ 1083, 1 ], [ 1083, 5 ], [ 1089, 3 ], [ 1092, 68 ], [ 1092, 69 ], [ 1101, 1 ],\\
& [ 1116, 11 ], [ 1119, 1 ], [ 1137, 1 ], [ 1143, 1 ], [ 1161, 6 ], [ 1161, 9 ], [ 1161, 10 ], [ 1161, 11 ], [ 1161, 12 ],\\
& [ 1164, 14 ], [ 1176, 57 ], [ 1176, 243 ], [ 1191, 1 ], [ 1197, 3 ], [ 1197, 4 ], [ 1200, 183 ], [ 1200, 384 ],\\
& [ 1200, 682 ], [ 1209, 3 ], [ 1209, 4 ], [ 1227, 1 ], [ 1236, 11 ], [ 1251, 1 ], [ 1263, 1 ], [ 1281, 3 ], [ 1281, 4 ],\\
& [ 1296, 3 ], [ 1296, 35 ], [ 1296, 37 ], [ 1296, 39 ], [ 1296, 220 ], [ 1296, 222 ], [ 1296, 226 ], [ 1296, 227 ],\\
& [ 1296, 228 ], [ 1296, 237 ], [ 1296, 647 ], [ 1296, 688 ], [ 1296, 689 ], [ 1296, 699 ], [ 1296, 1239 ], [ 1296, 1499 ],\\
& [ 1296, 1995 ], [ 1296, 2113 ], [ 1296, 2203 ], [ 1299, 1 ], [ 1308, 14 ], [ 1317, 1 ], [ 1323, 1 ], [ 1323, 4 ],\\
& [ 1323, 5 ], [ 1323, 7 ], [ 1323, 8 ], [ 1323, 14 ], [ 1323, 40 ], [ 1323, 42 ], [ 1323, 43 ], [ 1332, 14 ], [ 1344, 393 ],\\
& [ 1350, 46 ], [ 1359, 1 ], [ 1371, 1 ], [ 1389, 1 ], [ 1404, 14 ], [ 1404, 137 ], [ 1404, 138 ], [ 1404, 140 ],\\
& [ 1404, 141 ], [ 1407, 3 ], [ 1407, 4 ], [ 1413, 1 ], [ 1425, 5 ], [ 1443, 3 ], [ 1443, 4 ], [ 1452, 11 ], [ 1452, 34 ],\\
& [ 1458, 615 ], [ 1458, 618 ], [ 1458, 659 ], [ 1458, 663 ], [ 1458, 666 ], [ 1458, 1095 ], [ 1458, 1354 ], [ 1458, 1371 ],\\
& [ 1461, 1 ], [ 1467, 1 ], [ 1488, 57 ], [ 1497, 1 ], [ 1521, 1 ], [ 1521, 7 ], [ 1524, 11 ], [ 1533, 3 ], [ 1533, 4 ].
\end{split}
\end{equation}
\end{footnotesize}

\section{Relations between $G_\ell$ and $G_\nu$}\label{appB}

In this appendix we will show that for direct models, in order to obtain a phenomenologically
viable mixing matrix $U_\mathrm{PMNS}$, the generator $S$ of $G_\nu$ must not
commute with any of the generators of $G_\ell$. Note that we here exclude generators
of $G_\ell$ which are proportional to $\mathbbm{1}_3$, because such elements
do not restrict $U_\ell$ and are therefore superfluous.

Suppose a generator $T$ of $G_\ell$ commutes with $S$.
Since we have excluded generators $T$ which are proportional
to $\mathbbm{1}_3$, $T$ must have at least two different eigenvalues.
Therefore, $T$ alone fixes one column $u$ of $U_\ell$. Since $S$
commutes with $T$, $u$ is also an eigenvector of $S$.
However, since all eigenvectors of $S$ are unique (because $S$ fixes $U_\nu$),
$u$ is also a column of $U_\nu$. Therefore, $U_\ell$ and $U_\nu$
have two equal columns and $U_\mathrm{PMNS}$ is block-diagonal.
Thus two mixing angles vanish, which is phenomenologically excluded.

\section{The character table of the group $Q(648)$}

For completeness, we show the character table of the group $Q(648)=[648,533]$
constructed with GAP in table~\ref{charactertable}.

\begin{table}
\begin{center}
\resizebox{1.1\textwidth}{!}{
\begin{tabular}{|l|cccccccccccccccccccccccc|}
\hline
 & $1C^1$ & $1C^3$ & $1C^3$ & $24C^3$ & $9C^2$ & $9C^6$ & $9C^6$ & $54C^4$ & $54C^{12}$ & $54C^{12}$ & $12C^3$
& $12C^3$ & $12C^3$ & $72C^9$ & $36C^6$ & $36C^6$ & $36C^6$ & $12C^3$ & $12C^3$ & $12C^3$ & $72C^9$ & $36C^6$ & $36C^6$ & $36C^6$ \\
\hline
$\mathbf{1_{1}}$ & $1$ & $1$ & $1$ & $1$ & $1$ & $1$ & $1$ & $1$ & $1$ & $1$ & $1$ & $1$ & $1$ & $1$ & $1$ & $1$ & $1$ & $1$ & $1$ & $1$ & $1$ & $1$ & $1$ & $1$ \\
$\mathbf{1_{2}}$ & $1$ & $1$ & $1$ & $1$ & $1$ & $1$ & $1$ & $1$ & $1$ & $1$ & $\omega^2$ & $\omega^2$ & $\omega^2$ & $\omega^2$ & $\omega^2$ & $\omega^2$ & $\omega^2$ & $\omega$ & $\omega$ & $\omega$ & $\omega$ & $\omega$ & $\omega$ & $\omega$ \\
$\mathbf{1_{3}}$ & $1$ & $1$ & $1$ & $1$ & $1$ & $1$ & $1$ & $1$ & $1$ & $1$ & $\omega$ & $\omega$ & $\omega$ & $\omega$ & $\omega$ & $\omega$ & $\omega$ & $\omega^2$ & $\omega^2$ & $\omega^2$ & $\omega^2$ & $\omega^2$ & $\omega^2$ & $\omega^2$ \\
$\mathbf{2_{1}}$ & $2$ & $2$ & $2$ & $2$ & $-2$ & $-2$ & $-2$ & $0$ & $0$ & $0$ & $-1$ & $-1$ & $-1$ & $-1$ & $1$ & $1$ & $1$ & $-1$ & $-1$ & $-1$ & $-1$ & $1$ & $1$ & $1$ \\
$\mathbf{2_{2}}$ & $2$ & $2$ & $2$ & $2$ & $-2$ & $-2$ & $-2$ & $0$ & $0$ & $0$ & $-\omega$ & $-\omega$ & $-\omega$ & $-\omega$ & $\omega$ & $\omega$ & $\omega$ & $-\omega^2$ & $-\omega^2$ & $-\omega^2$ & $-\omega^2$ & $\omega^2$ & $\omega^2$ & $\omega^2$ \\
$\mathbf{2_{3}}$ & $2$ & $2$ & $2$ & $2$ & $-2$ & $-2$ & $-2$ & $0$ & $0$ & $0$ & $-\omega^2$ & $-\omega^2$ & $-\omega^2$ & $-\omega^2$ & $\omega^2$ & $\omega^2$ & $\omega^2$ & $-\omega$ & $-\omega$ & $-\omega$ & $-\omega$ & $\omega$ & $\omega$ & $\omega$ \\
$\mathbf{3_{1}}$ & $3$ & $3$ & $3$ & $3$ & $3$ & $3$ & $3$ & $-1$ & $-1$ & $-1$ & $0$ & $0$ & $0$ & $0$ & $0$ & $0$ & $0$ & $0$ & $0$ & $0$ & $0$ & $0$ & $0$ & $0$ \\
$\mathbf{3_{2}}$ & $3$ & $3 \omega^2$ & $3 \omega$ & $0$ & $-1$ & $-\omega^2$ & $-\omega$ & $1$ & $\omega^2$ & $\omega$ & $-2 \omega-\omega^2$ & $\omega+2 \omega^2$ & $\omega-\omega^2$ & $0$ & $-\omega^2$ & $-\omega$ & $-1$ & $-\omega-2 \omega^2$ & $-\omega+\omega^2$ & $2 \omega+\omega^2$ & $0$ & $-\omega$ & $-1$ & $-\omega^2$ \\
$\mathbf{3_{3}}$ & $3$ & $3 \omega$ & $3 \omega^2$ & $0$ & $-1$ & $-\omega$ & $-\omega^2$ & $1$ & $\omega$ & $\omega^2$ & $-\omega-2 \omega^2$ & $2 \omega+\omega^2$ & $-\omega+\omega^2$ & $0$ & $-\omega$ & $-\omega^2$ & $-1$ & $-2 \omega-\omega^2$ & $\omega-\omega^2$ & $\omega+2 \omega^2$ & $0$ & $-\omega^2$ & $-1$ & $-\omega$ \\
$\mathbf{3_{4}}$ & $3$ & $3 \omega^2$ & $3 \omega$ & $0$ & $-1$ & $-\omega^2$ & $-\omega$ & $1$ & $\omega^2$ & $\omega$ & $\omega+2 \omega^2$ & $\omega-\omega^2$ & $-2 \omega-\omega^2$ & $0$ & $-\omega$ & $-1$ & $-\omega^2$ & $2 \omega+\omega^2$ & $-\omega-2 \omega^2$ & $-\omega+\omega^2$ & $0$ & $-\omega^2$ & $-\omega$ & $-1$ \\
$\mathbf{3_{5}}$ & $3$ & $3 \omega$ & $3 \omega^2$ & $0$ & $-1$ & $-\omega$ & $-\omega^2$ & $1$ & $\omega$ & $\omega^2$ & $2 \omega+\omega^2$ & $-\omega+\omega^2$ & $-\omega-2 \omega^2$ & $0$ & $-\omega^2$ & $-1$ & $-\omega$ & $\omega+2 \omega^2$ & $-2 \omega-\omega^2$ & $\omega-\omega^2$ & $0$ & $-\omega$ & $-\omega^2$ & $-1$ \\
$\mathbf{3_{6}}$ & $3$ & $3 \omega^2$ & $3 \omega$ & $0$ & $-1$ & $-\omega^2$ & $-\omega$ & $1$ & $\omega^2$ & $\omega$ & $\omega-\omega^2$ & $-2 \omega-\omega^2$ & $\omega+2 \omega^2$ & $0$ & $-1$ & $-\omega^2$ & $-\omega$ & $-\omega+\omega^2$ & $2 \omega+\omega^2$ & $-\omega-2 \omega^2$ & $0$ & $-1$ & $-\omega^2$ & $-\omega$ \\
$\mathbf{3_{7}}$ & $3$ & $3 \omega$ & $3 \omega^2$ & $0$ & $-1$ & $-\omega$ & $-\omega^2$ & $1$ & $\omega$ & $\omega^2$ & $-\omega+\omega^2$ & $-\omega-2 \omega^2$ & $2 \omega+\omega^2$ & $0$ & $-1$ & $-\omega$ & $-\omega^2$ & $\omega-\omega^2$ & $\omega+2 \omega^2$ & $-2 \omega-\omega^2$ & $0$ & $-1$ & $-\omega$ & $-\omega^2$ \\
$\mathbf{6_{1}}$ & $6$ & $6 \omega^2$ & $6 \omega$ & $0$ & $2$ & $2 \omega^2$ & $2 \omega$ & $0$ & $0$ & $0$ & $-\omega+\omega^2$ & $2 \omega+\omega^2$ & $-\omega-2 \omega^2$ & $0$ & $-1$ & $-\omega^2$ & $-\omega$ & $\omega-\omega^2$ & $-2 \omega-\omega^2$ & $\omega+2 \omega^2$ & $0$ & $-1$ & $-\omega^2$ & $-\omega$ \\
$\mathbf{6_{2}}$ & $6$ & $6 \omega$ & $6 \omega^2$ & $0$ & $2$ & $2 \omega$ & $2 \omega^2$ & $0$ & $0$ & $0$ & $\omega-\omega^2$ & $\omega+2 \omega^2$ & $-2 \omega-\omega^2$ & $0$ & $-1$ & $-\omega$ & $-\omega^2$ & $-\omega+\omega^2$ & $-\omega-2 \omega^2$ & $2 \omega+\omega^2$ & $0$ & $-1$ & $-\omega$ & $-\omega^2$ \\
$\mathbf{6_{3}}$ & $6$ & $6 \omega^2$ & $6 \omega$ & $0$ & $2$ & $2 \omega^2$ & $2 \omega$ & $0$ & $0$ & $0$ & $2 \omega+\omega^2$ & $-\omega-2 \omega^2$ & $-\omega+\omega^2$ & $0$ & $-\omega^2$ & $-\omega$ & $-1$ & $\omega+2 \omega^2$ & $\omega-\omega^2$ & $-2 \omega-\omega^2$ & $0$ & $-\omega$ & $-1$ & $-\omega^2$ \\
$\mathbf{6_{4}}$ & $6$ & $6 \omega$ & $6 \omega^2$ & $0$ & $2$ & $2 \omega$ & $2 \omega^2$ & $0$ & $0$ & $0$ & $\omega+2 \omega^2$ & $-2 \omega-\omega^2$ & $\omega-\omega^2$ & $0$ & $-\omega$ & $-\omega^2$ & $-1$ & $2 \omega+\omega^2$ & $-\omega+\omega^2$ & $-\omega-2 \omega^2$ & $0$ & $-\omega^2$ & $-1$ & $-\omega$ \\
$\mathbf{6_{5}}$ & $6$ & $6 \omega^2$ & $6 \omega$ & $0$ & $2$ & $2 \omega^2$ & $2 \omega$ & $0$ & $0$ & $0$ & $-\omega-2 \omega^2$ & $-\omega+\omega^2$ & $2 \omega+\omega^2$ & $0$ & $-\omega$ & $-1$ & $-\omega^2$ & $-2 \omega-\omega^2$ & $\omega+2 \omega^2$ & $\omega-\omega^2$ & $0$ & $-\omega^2$ & $-\omega$ & $-1$ \\
$\mathbf{6_{6}}$ & $6$ & $6 \omega$ & $6 \omega^2$ & $0$ & $2$ & $2 \omega$ & $2 \omega^2$ & $0$ & $0$ & $0$ & $-2 \omega-\omega^2$ & $\omega-\omega^2$ & $\omega+2 \omega^2$ & $0$ & $-\omega^2$ & $-1$ & $-\omega$ & $-\omega-2 \omega^2$ & $2 \omega+\omega^2$ & $-\omega+\omega^2$ & $0$ & $-\omega$ & $-\omega^2$ & $-1$ \\
$\mathbf{8_{1}}$ & $8$ & $8$ & $8$ & $-1$ & $0$ & $0$ & $0$ & $0$ & $0$ & $0$ & $2$ & $2$ & $2$ & $-1$ & $0$ & $0$ & $0$ & $2$ & $2$ & $2$ & $-1$ & $0$ & $0$ & $0$ \\
$\mathbf{8_{2}}$ & $8$ & $8$ & $8$ & $-1$ & $0$ & $0$ & $0$ & $0$ & $0$ & $0$ & $2 \omega^2$ & $2 \omega^2$ & $2 \omega^2$ & $-\omega^2$ & $0$ & $0$ & $0$ & $2 \omega$ & $2 \omega$ & $2 \omega$ & $-\omega$ & $0$ & $0$ & $0$ \\
$\mathbf{8_{3}}$ & $8$ & $8$ & $8$ & $-1$ & $0$ & $0$ & $0$ & $0$ & $0$ & $0$ & $2 \omega$ & $2 \omega$ & $2 \omega$ & $-\omega$ & $0$ & $0$ & $0$ & $2 \omega^2$ & $2 \omega^2$ & $2 \omega^2$ & $-\omega^2$ & $0$ & $0$ & $0$ \\
$\mathbf{9_{1}}$ & $9$ & $9 \omega^2$ & $9 \omega$ & $0$ & $-3$ & $-3 \omega^2$ & $-3 \omega$ & $-1$ & $-\omega^2$ & $-\omega$ & $0$ & $0$ & $0$ & $0$ & $0$ & $0$ & $0$ & $0$ & $0$ & $0$ & $0$ & $0$ & $0$ & $0$ \\
$\mathbf{9_{2}}$ & $9$ & $9 \omega$ & $9 \omega^2$ & $0$ & $-3$ & $-3 \omega$ & $-3 \omega^2$ & $-1$ & $-\omega$ & $-\omega^2$ & $0$ & $0$ & $0$ & $0$ & $0$ & $0$ & $0$ & $0$ & $0$ & $0$ & $0$ & $0$ & $0$ & $0$ \\
\hline
\end{tabular}
}
\caption{The character table of the group $Q(648)=[648,533]$ constructed with GAP. The notation for the conjugacy classes
of the group is $N C^{n}$, where $N$ denotes the number of elements of the conjugacy class
and $n$ is the order of the elements of the class. ($\omega=\exp(2\pi i/3).$)\label{charactertable}}
\end{center}
\end{table}

\end{appendix}

\end{document}